\documentclass[letterpaper, 10 pt, conference]{ieeeconf}  % Comment this line out
\IEEEoverridecommandlockouts                              % This command is only
\usepackage{graphicx} % for pdf, bitmapped graphics files

\title{\LARGE \bf
Limitations of Quantum Advantage in Unsupervised Machine Learning
}

\author{Apoorva D. Patel% <-this % stops a space
\thanks{A.D. Patel is with the Centre for High Energy Physics, Indian Institute of Science, Bangalore 560012, India, and the International Centre for Theoretical Sciences, Bangalore 560089, India. {\tt\small adpatel@iisc.ac.in}}%
}

\begin{document}

\maketitle
\thispagestyle{empty}
\pagestyle{empty}

%%%%%%%%%%%%%%%%%%%%%%%%%%%%%%%%%%%%%%%%%%%%%%%%%%%%%%%%%%%%%%%%%%%%%%%%%%%%%%%%
\begin{abstract}

Machine learning models are used for pattern recognition analysis of big data,
without direct human intervention. The task of unsupervised learning is to
find the probability distribution that would best describe the available data,
and then use it to make predictions for observables of interest. Classical
models generally fit the data to Boltzmann distribution of Hamiltonians with
a large number of tunable parameters. Quantum extensions of these models
replace classical probability distributions with quantum density matrices.
An advantage can be obtained only when features of density matrices that are
absent in classical probability distributions are exploited. Such situations
depend on the input data as well as the targeted observables. Explicit
examples are discussed that bring out the constraints limiting possible
quantum advantage. The problem-dependent extent of quantum advantage has
implications for both data analysis and sensing applications.

\end{abstract}

%%%%%%%%%%%%%%%%%%%%%%%%%%%%%%%%%%%%%%%%%%%%%%%%%%%%%%%%%%%%%%%%%%%%%%%%%%%%%%%%
\section{BACKGROUND}

Over the last decade, the subject of machine learning has grown explosively
in many directions. One reason for that is the availability of big data in
many fields. Various types of sensors and detectors have become affordable
and convenient. They collect huge amount of data (with space and time labels),
which need to be analysed to make decisions. Often there is no time and space
to store the data; interesting features must be extracted quickly and the
remaining data discarded. The challenge is then to carry out these tasks
efficiently. Examples of such situations span astronomy, satellite imagery,
surveillance, weather and climate, internet traffic, model system tests and
validation (for design purposes), experimental observations, collider physics
and genetic information.

The analysis tackling these problems typically undergoes three steps:
recognise interesting patterns, classify them and then make appropriate
recommendations. The tedious job of learning from the big data is given to
a computer, together with some rules, so that the results can be extracted
without much human intervention. The popular learning algorithms based on a
large number of tunable parameters include \cite{Kelleher,Russell,Sutton}:\\
$\bullet$ Supervised learning, where training data with known pattern labels
is provided to the computer program.\\
$\bullet$ Unsupervised learning, where similar examples and analogies are
provided to the computer program.\\
$\bullet$ Reinforcement learning, where reward functions evaluating the
extracted results are provided to the computer program.

These algorithms generally use neural network methods, heuristically and
without proof, and obtain probabilistic results. Moreover, the analysis uses
fast linear algebra techniques. When the number of tunable parameters is
small, say tens or hundreds, the results do not offer much predictive power.
But when the number of tunable parameters becomes large, going up to millions
and billions, there is a vast improvement in predictive power. Affordable
availability of computational resources that can handle such large number
of tunable parameters, with multiple GPU systems, is the other reason for
the rapid advances in this field.

Almost parallel developments in the field of quantum computation has added
another aspect to the adventure; may be one can achieve more by a suitable
combination of classical and quantum strategies \cite{Schuld}. The aim of
this presentation is to quantify the extra advantage that can be obtained
by combination of the two strategies, in the specific case of unsupervised
learning.

\section{UNSUPERVISED LEARNING}

The task of unsupervised learning is to ascertain the probability distribution
of the data generated in a specific scenario, which can then be used to
make predictions. Initially one guesses the probability distribution based
on analogies, and then systematically evolves it to improve the approximation.
This strategy needs two ingredients: a criterion to measure the difference
between two probability distributions, and an algorithm to reduce the
difference by varying parameters characterising the guessed distribution.

A good criterion to measure the difference between two probability
distributions is the Kullback-Leibler (KL) divergence, i.e. the relative
entropy of the two distributions:
\begin{equation}
D_{KL}(p\|q) = \sum_x p(x) \log(p(x)/q(x)) ~,
\end{equation}
where $x$ denotes the data labels. It is the amount of information lost
when the distribution $p(x)$ is used to approximate the distribution $q(x)$.
It is asymmetric in $p(x)$ and $q(x)$, but it is always non-negative and
vanishes only when $p(x)$ and $q(x)$ coincide \cite{Nielsen}. So a
parametrised $p(x)$ is fit to the data $q(x)$, by minimising $D_{KL}(p\|q)$.

When $D_{KL}(p\|q)$ is expanded around its minimum, which is zero, the leading
quadratic form gives the Fisher information metric, whose eigendirections
obey the corresponding Cram\'er-Rao bounds for the estimation of parameters.

\subsection{Boltzmann Machines}

Of course, one cannot make much progress without a practical parametrisation
of $p(x)$. The crucial assumption that is made is that the target probability
distribution $q(x)$ arose from some physical process, and so it can be
reasonably approximated by the thermal Boltzmann distribution,
\begin{equation}
p(x) = e^{-\beta H(x)}/(\sum_{states~x} e^{-\beta H(x)}) ~,
\end{equation}
in terms of a local Hamiltonian $H(x)$ containing many variational parameters.
A neural network parametrisation of these distributions is called a Boltzmann
machine, with the state variables $x$ represented by network nodes.

In this parametrisation, when some degrees of freedom are not observed,
they get summed over. Such a sum produces reduced probability distributions;
the unobserved degrees of freedom are mapped to hidden variables of a neural
network, while the rest remain as the observable (or visible) degrees of
freedom.

The favoured realisations of Boltzmann machines are stochastic Ising
spin-glass models with an external field. They are versatile enough to
describe a variety of physical behaviour (e.g. regular, periodic, glassy,
chaotic), as well as are easy to simulate. They are defined by the Hamiltonian
\begin{equation}
H = - \sum_{ij} w_{ij} s_i s_j - \sum_i b_i s_i ~,
\end{equation}
where $w_{ij}$ is the interaction strength between the spins $s_i$ and $s_j$,
and $b_i$ is the external bias field. In simulations starting from a high
temperature, the spin network evolves from a random configuration towards
its thermal equilibrium state (i.e. simulated annealing).

The Hamiltonian defined above has too many parameters, so based on observed
interaction structure between neurons, it is simplified to construct
restricted Boltzmann machines where interactions exist only between the
observable and the hidden variables but not among the observable ones
or among the hidden ones. Such a formulation is further extended to define
deep Boltzmann machines with multiple layers of hidden variables, where
interactions exist between successive layers but not within any particular
layer. Fig. 1 illustrates such a network.

\begin{figure}[b]
\centering
\includegraphics[scale=0.8]{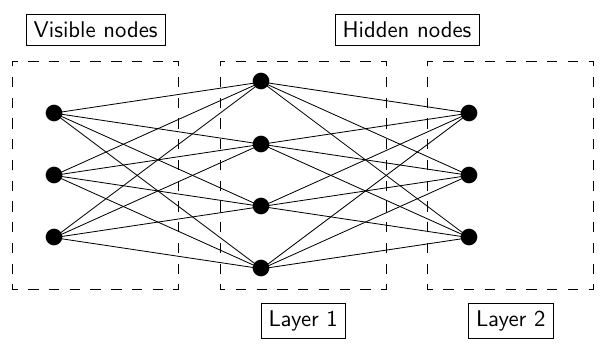}
\caption{An Example of a deep Boltzmann machine with one visible and two
hidden layers. The lines denote the interactions between nodes.}
\label{NeuralNetFig}
\end{figure}

\section{QUANTUM GENERALISATION}

In the quantum setting, the classical probability distribution generalises
to the quantum density matrix $\rho(x)$, which provides the complete
description of any quantum state. The expectation values of observables
undergo the replacement: 
$\langle O\rangle = \sum_x p(x)O(x) \longrightarrow Tr[\rho(x) O(x)]$.
Since $\rho(x)$ is positive semi-definite, it can always be expressed as
$e^{-\beta H(x)}$ in terms of a suitable quantum Hamiltonian.

With this change, the reduced probability distributions generalise to
reduced density matrices obtained by partial trace over hidden variables.
The relative entropy of quantum distributions generalises to
\begin{equation}
S(\rho\|\sigma) = Tr[\rho(x)(\log\rho(x) - \log\sigma(x))] ~,
\end{equation}
which remains non-negative and vanishes only when $\rho(x)$ and $\sigma(x)$
coincide \cite{Nielsen}. Also, the relation between $S(\rho\|\sigma)$
and the quantum Fisher information metric (and the Cram\'er-Rao bounds)
has the same structure as in the classical case.

The novel quantum feature is that the Hermitian objects $\rho(x)$ and $O(x)$
may not commute. When they do, they can be simultaneously diagonalised,
$Tr[\rho(x) O(x)]$ simplifies to $\sum_x p(x)O(x)$, and the whole quantum
analysis reduces to the classical case. Thus $[\rho,O]\ne0$ is a necessary
condition for obtaining quantum advantage in any unsupervised learning problem.
It is worthwhile to note that the quantum advantage depends on the observable
chosen for a given $\rho(x)$, while the same $p(x)$ applies to all observables
in the classical analysis.

Turning the argument around, the quantum advantage is expected to be the
maximum when the norm of $[\rho,O]$ is the largest. With $\|\rho\| \le 1$,
and $\|O\|$ fixed by some convention, this is an explicitly solvable exercise.

\subsection{Quantum Advantage}

With unitary transformations, $\rho$ and $O$ can be diagonalised in their own
eigenbases. Then, $\langle O\rangle = Tr[\rho_{diag} U O_{diag} U^\dagger]$,
where $U$ is the relative unitary rotation between the two eigenbases. This
expression can be interpreted to solve two types of problems: What is the
optimal probe state $\rho$ that gives maximum quantum advantage in the
determination of an observable $O$, and which observables $O$ can be
determined with maximum quantum sensitivity given a prepared state $\rho$?

The example of a single qubit illustrates the answers clearly. The generic
qubit density matrix is $\rho=\frac{1}{2}(I+\vec{r}\cdot\vec{\sigma})$ with
$|\vec{r}| \le 1$, which can be geometrically represented on the Bloch
sphere. For an $O$ diagonal in the $Z$-basis, the norm of $[\rho,O]$ is
the maximum when $\vec{r}$ is in the equatorial plane, and $|\vec{r}|=1$
is the best choice. Starting from initial $\rho$ along the $Z$-axis, such
a $\rho$ can be prepared by rotating it about an axis in the equatorial
plane by angle $\frac{\pi}{2}$.

Astronomers have been using this property to accurately detect the
polarisation of electromagnetic wave signals; the filters are oriented
so as to be transverse to the polarisation direction to be measured.
The technique has also been extended to design appropriate geometrical
setting for accurate determination of gravitational wave signals; the two
arms of the interferometer are oriented orthogonal to each other for the
best identification of the quadrupolar gravitational radiation.

% For a generic mixed state $\rho$, and any Pauli-basis operator $O$,
% $Tr(\rho O)$ can be easily determined: Initialise a probe qubit in the
% state $|+\rangle$, use it to perform a ctrl-$O$ operation on $\rho$,
% and then measure it in the $X$-basis.

\subsection{Maximising Quantum Advantage}

A naive multi-qubit generalisation of the previous single qubit optimisation
analysis is provided by the transverse field quantum Heisenberg model:
\begin{equation} H = - \sum_{ij} w_{ij} \sigma_i^{(z)} \sigma_j^{(z)}
- \sum_i b_i \sigma_i^{(x)} ~.
\end{equation}
Attempts have been made to iteratively optimise its parameters in order to
maximise quantum advantage \cite{Amin}. The maximisation problem can be
solved explicitly in any dimension, however, as described in what follows.

Note that only the traceless parts of $\rho$ and $O$ contribute to $[\rho,O]$.
Now choose the basis in which $O$ is diagonal. In an $n$-dimensional Hilbert
space, its traceless part can be expressed as a linear combination of $n-1$
independent and orthogonal Cartan generators. If only one such generator is
present in $O$, of the form $diag(\ldots,1,\ldots,-1,\ldots)$, then similar
to the case of a single qubit, the norm $\| [\rho,O] \|$ is maximised for
the two-dimensional pure state
\begin{equation}
\rho = diag(\ldots,\frac{1}{2},\ldots,\frac{1}{2},\ldots)
     + c~\vec{\sigma}\cdot\hat{n}_T ~.
\end{equation}
Here $\hat{n}_T$ in the direction transverse to the generator,
and $c=\frac{1}{2}$.

When multiple Cartan generators contribute to the traceless part of $O$,
its diagonal structure is of the form:
$diag(a_1, \ldots,a_p,-b_1,\ldots,-b_q)$, $p+q=n$, $a_i,b_j\ge0$,
$\sum_i a_i=t=\sum_j b_j$.
Then the norm $\| [\rho,O] \|$ is maximised by constructing the two-component
generator for $O$, $T=diag(\ldots,t,\ldots,-t,\ldots)$, where $t$ is in one
of the $p$ directions, $-t$ is in one of the $q$ directions and the rest of
the components are zero. (This choice is highly degenerate.)
All the remaining traceless contributions to $O$ orthogonal to $T$ have
the form $diag(\ldots,1,\ldots,1,\ldots)$, where $1$'s are in the same
locations as $\pm t$, the other components are related to $a_i$ and $b_j$,
and each such contribution has magnitude less than $t$. The maximum value
for the norm $\| [\rho,O] \|$ is then obtained by aligning $\rho$ with the
Cartan contribution of the largest magnitude, $T$. (Strictly speaking, this
choice maximises the infinity norm $\| [\rho,O] \|_\infty$.) The result is
again the transverse direction choice specified in (6).

Furthermore, it is useful to choose $T$ such that the locations of $t$ and
$-t$ match those of $max(a_i)$ and $max(b_j)$. That maximises the spectral
gap of the Cartan contributions, speeding up the convergence of the norm
$\| [\rho,O] \|$ to its maximum.

Such a $\rho$ aligned with $T$ can be efficiently prepared. Starting from
an $m$-qubit register in the zero-state (choosing $m=\lceil\log_2n\rceil$
can pad up the density matrix with extra non-contributing states), first
one can transform to the pure state in the direction given by $max(a_i)$,
using $O(m)$ operations. Then this state can be aligned with $\hat{n}_T$,
by rotation about the equatorial axis orthogonal to $\hat{n}_T$, in the
subspace specified by $max(a_i)$ and $max(b_j)$. Performing this subspace
rotation also takes $O(m)$ operations. The freedom to reorder the Cartan
generator components of $O$ by permutations can actually help simplify
this preparation. For example, the locations of $max(a_i)$ and $max(b_j)$
can be assigned the least significant bit of the $m$-qubit register,
and then the desired $\rho$ can be prepared by operations on the least
significant qubit, just as in the single qubit case described in the
previous subsection.

\section{LESSONS}

The analysis of the previous section has powerful implications that
constrain the quantum advantage of Boltzmann machines:\\
$\bullet$ Possible quantum advantage of a Boltzmann machine is
problem-dependent, i.e. it depends on the relative orientation of $\rho$
and $O$. Analytical maximisation of $\| [\rho,O] \|$ skips the need to
iteratively minimise $S(\rho\|\sigma)$.\\
$\bullet$ The quantum advantage is maximum when $\rho$ is a pure state.
This is expected because interaction with the environment generically
reduces quantum effects. But it also implies that the visible state maximising
quantum advantage is fully decoupled from all the hidden degrees of freedom.
The hidden degrees of freedom can then provide only classical advantage.\\
$\bullet$ Even when quantum correlations are allowed to spread from the
visible to the hidden degrees of freedom, the spread is limited as per the
Schmidt decomposition, and the quantum advantage offered by the mixed state
$\rho_v$ is less than that for a pure state $\rho_v$. The $m$ visible qubits
can couple to at most $m$ configurable hidden qubits,
$|\psi\rangle_{vh} = \sum_i \sqrt{p_i} |i\rangle_v \otimes |i'\rangle_h$ in
the basis that diagonalises the reduced $\rho_v$. The quantum correlations
cannot extend beyond this limited range, unlike classical correlations,
and any advantage offered by Boltzmann machines by adding more hidden
degrees of freedom is completely classical. The one-to-one map of the
Schmidt decomposition also implies that restricted Boltzmann machines
suffice for getting any quantum advantage.\\
$\bullet$ For the measurement of a specific $O$, the optimal transverse
$\rho$ offers the best sensitivity as per the Cram\'er-Rao bound. It has
been shown that the optimal quantum state to sense a quantum signal is the
eigenvector corresponding to the largest eigenvalue of the quantum Fisher
information metric \cite{Reilly}, and we have determined the explicit
solution for the eigenvector here.

In general, when looking for quantum advantage, one must remember that the
quantum theory originated with introduction of a new fundamental constant
of nature, the Planck constant. It is the commutator of canonically conjugate
pair of coordinates, is the minimal area in the structure of the phase space,
is the unit of angular momentum, and part of the definition of the energy
quantum. Quantum advantage is possible only when features absent in classical
theory are present in a quantum algorithm, and the involvement of a nonzero
Planck constant is indispensable for that to happen. The necessity of a
nonzero $[\rho,O]$ to obtain quantum advantage in unsupervised learning,
discussed in this article, is a specific example of this constraint.
Another well-known example illustrating this constraint is the von Neumann
non-demolition measurement, where the variable canonically conjugate to the
observed variable (it is the generator for translation of the observed
variable) governs the system-detector interaction Hamiltonian \cite{Neumann}.

The constraints on unsupervised learning obtained here can be relaxed,
when the input data are quantum in nature (the output is always classical).
When the input data originate from a quantum process, their multiple
correlated streams can be directly fed to a quantum processor, without
intervening measurements that project them to classical data. Retaining
coherent correlations between even two parallel data streams (encoded as
density matrices) can provide an exponential quantum advantage \cite{Huang}.

%%%%%%%%%%%%%%%%%%%%%%%%%%%%%%%%%%%%%%%%%%%%%%%%%%%%%%%%%%%%%%%%%%%%%%%%%%%%%%%%
\addtolength{\textheight}{-3cm}   % This command serves to balance the column lengths

\section{ACKNOWLEDGMENTS}

This work was supported in part by the Accenture Project ``Quantum Boltzmann
Machines" (PC53014) under SID at IISc, and the research project ``Quantum
Algorithms and Simulations" under DIA-RCoE at IISc.


\begin{thebibliography}{99}

\bibitem{Kelleher} J.~D. Kelleher, B. Mac Namee and A. D'Arcy,
        {\it Fundamentals of Machine Learning for Predictive Data Analytics:
        Algorithms, Worked Examples, and Case Studies}, Second Edition,
        The MIT Press, Cambridge MA, USA; 2015.
\bibitem{Russell} S.~J. Russell and P. Norvig,
        {\it Artificial Intelligence: A Modern Approach}, Fourth Edition,
        Pearson, India; 2022.
\bibitem{Sutton} R.~S. Sutton and A.~G. Barto,
        {\it Reinforcement Learning: An Introduction}, Second Edition,
        The MIT Press, Cambridge MA, USA; 2020.
\bibitem{Schuld} M. Schuld and F. Petruccione,
        {\it Machine Learning with Quantum Computers}, Second Edition,
        Springer Nature Switzerland AG; 2021.
\bibitem{Nielsen} See for instance: M.~A. Nielsen and I.~L. Chuang,
        {\it Quantum Computation and Quantum Information},
        Cambridge University Press, Cambridge, UK; 2000.
\bibitem{Amin} See for instance: M.~H. Amin, E. Andriyash,, J. Rolfe,
        B. Kulchytskyy and R. Melko, {\it Phys. Rev. X} 8, 2018,
        article 021050.
\bibitem{Reilly} J.~T. Reilly, J.~D. Wilson, S.~B. J\"ager, C. Wilson and
        M.~J. Holland, {\it Phys. Rev. Lett.} 131, 2023, article 150802.
\bibitem{Neumann} J. von Neumann, 
        {\it Mathematical Foundations of Quantum Mechanics}, New Edition,
        Princeton University Press, USA; 1955.
\bibitem{Huang} H.-Y. Huang, R. Kueng and J. Preskill, {\it Phys. Rev. Lett.}
        126, 2021, article 190505.

\end{thebibliography}
\end{document}